# "Phonon lasing" as a likely mechanism for density-dependent velocity saturation in GaN transistors.

Jacob B. Khurgin[1], Sanyam Bajaj[2] and Siddharth Rajan[2]

[1]Department of Electrical & Computer Engineering, Johns Hopkins University, Baltimore, MD 21218

[2]Department of Electrical & Computer Engineering, The Ohio State University, Columbus, OH 43210

We show that density-dependent velocity saturation in a GaN High Electron Mobility Transistor (HEMT) can be related to the stimulated emission of longitudinal optical (LO) phonons. As the drift velocity of electrons increases, the drift of the Fermi distribution in reciprocal space produces population inversion and gain for the LO phonons. Once this gain reaches a threshold value, the avalanche-like increase of LO emission causes a rapid loss of electron energy and momentum and leads to drift velocity saturation. Our simple model correctly predicts both the general trend of the saturation velocity decreasing with increasing electron density and the values of saturation velocity measured in our experiments.

HEMTs based on GaN and other group III Nitrides are among the most promising devices for high power amplification in RF, extending all the way to hundreds of GHz [1-4]. GaN is a wide bandgap material with an insignificant non-parabolicity and with a satellite valley that is removed from the conduction band edge by as much as 1.4eV. In view of that, neither the non-parabolicity nor the inter-valley transfer can cause velocity saturation. At the same time, GaN is a polar material with strong coupling between the longitudinal lattice oscillations (LO phonons) and electrons and it is expected that, once the kinetic energy of electrons moving with the drift velocity $v_d$, $m_c v_d^2 / 2$ approaches the energy of the optical phonon $\hbar\omega_{LO} \approx 92 meV$, LO phonon emission should cause saturation of the velocity, roughly at somewhat less than $v_{max} \approx 4\times10^7 cm/s$, which is larger than the saturation velocities in narrow bandgap III-V semiconductors with much lower effective masses [5]. However, experimental data indicate that drift velocity saturates at lower values, in particular when the carrier concentration in the HEMT channel is high [1-4]. Recently we carried out a rigorous experimental study of velocity saturation in the GaN channel as a function of carrier concentration [6] and have found that saturation velocity $v_{sat}$ steadily decreases from about 1.9×10⁷ cm/s to less than $1.0\times10^7$ cm/s as the sheet density $n_s$ of carriers in the channel increases from $0.8\times10^{12}$ cm$^{-2}$ to about $8.0\times10^{12}$ cm$^{-2}$. High power operation of a HEMT requires high carrier density; hence it is critical to understand the origin of the density dependence of the velocity saturation, and, based on the knowledge gained, develop means of mitigating the factors that cause the saturation velocity to plunge at high carrier densities.

In the course of the last decade the consensus has been formed that velocity saturation in III-Nitrides is caused by hot longitudinal optical (LO) phonons which are generated by electron-phonon scattering and remain in the channel due to their low velocity and relatively long lifetime $\tau_p$ [7-11]. The latter can be as long as a few picoseconds [12,13] because the energy of an LO phonon is so high that it cannot efficiently decay into two acoustic phonons (so-called Klemens process [14]) and instead decays into a combination of transverse optical (TO) and acoustic phonons (so-called Ridley process [15], which is less efficient). It should be noted that indirect measurement of LO phonon lifetime using Anti-Stokes Raman [16] and microwave noise [17,18] measurements indicated the reduction of $\tau_p$ with electron density $n_s$, but in our recent publication [19] we have shown that at high carrier densities LO phonons only get scattered away from the center of Brillouin zone (BZ), while still remaining in the channel for a few picoseconds. While hot phonons are recognized as the source of velocity saturation in III-Nitrides, the nature of the exact mechanism causing the saturation remains unsettled. In general, it is assumed that since electrons are in thermal equilibrium with phonons, as the occupational numbers of LO phonons $n_{LO}$ increase, so do the electron temperature $T_e$ and momentum scattering rate $\tau_m^{-1} \sim (2n_{LO}+1)$, while the energy relaxation rate $\tau_E^{-1}$ stays roughly the same, which causes gradual velocity saturation. This follows from the fact that both emission and absorption of phonons cause change in the momentum, while the energy transfer is determined by the net difference between the emission and absorption of phonons. Obviously, $n_{LO}$ increases roughly proportionally to the 3-D density of carriers $n_e$; hence, as shown in [8], the saturation velocity should decrease as roughly $v_{sat} \sim n_e^{-1/2}$. The measurements performed in [6] agree qualitatively with the theory developed in [8] but cannot be approximated by the simple power dependence over the entire range of electron densities. Furthermore, the theory [8] predicts a rather gradual, "soft" velocity vs field dependence, rather than a hard clamping of velocity observed experimentally. One way to determine saturation velocity would be to perform full Monte-Carlo simulations [20], which are time consuming, depend on large number of input parameters, and often do not reveal the clear physical picture behind the observed effects. Thus a simple model capable of providing a realistic estimate of saturation velocity and its dependence on carrier density would be a welcome addition to the existing means of predicting performance of III-N devices. In this letter we propose such a model in which the saturation of velocity occurs when the probability of stimulated emission of LO phonons exceeds a certain threshold value determined by the LO phonon lifetime. Using this model and a very limited number of device/material parameters, we succeed in predicting the values of $v_{sat}(n_s)$ that are very close to the data obtained in [5]. The "clamping" of drift velocity above threshold is strikingly similar to the clamping of population inversion in lasers; hence one may refer to the mechanism responsible for the velocity saturation as "phonon lasing" [21,22].

In prior works [7,8] momentum and energy relaxation rates have always been estimated under the assumption of equilibrium distribution of electrons in the conduction band described by the Fermi-Dirac (FD) function $f_0(\mathbf{k}) = [\exp((k^2 - \mu)/T_e) + 1]^{-1}$, shown in Fig.1a, where all the variables have been normalized: energies including chemical potential $\mu$ to LO phonon energy $\hbar\omega_{LO}$, electron temperature $T_e$ to $T_0 = \hbar\omega_{LO}/k_B = 1060K$, and the wavevector $k$ to $q_0 = (2m_0\omega_{LO}/\hbar)^{1/2} = 0.68 nm^{-1}$. If one now consider the scattering by LO phonons between the state $\mathbf{k}_1$ with energy $E_1$ and state $\mathbf{k}_2$ with energy $E_2 = E_1 - \hbar\omega_{LO}$, the net probability of emitting an LO phonon with wavevector $\mathbf{q} = \mathbf{k}_1 - \mathbf{k}_2$ can be written as $R_{12} = 2\pi\hbar^{-1}M^2(R_{sp} + R_{stim})$, where $R_{sp} = f_0(\mathbf{k}_1)[1 - f_0(\mathbf{k}_2)]$ is the spontaneous emission term, $R_{st} = [f_0(\mathbf{k}_1) - f_0(\mathbf{k}_2)]n_q$ is the stimulated emission term, $n_q$ is the occupational number of LO phonons, $M = q^{-1}(e^2\hbar^2/\varepsilon' 4m_c V)^{1/2}$ is the matrix element of electron-phonon interaction, $1/\varepsilon' = 1/\varepsilon_\infty - 1/\varepsilon$, and $\varepsilon, \varepsilon_\infty$ are values of static and infrared dielectric permittivity respectively. Since state 1 has higher energy than state 2, $f_0(\mathbf{k}_1) < f_0(\mathbf{k}_2)$ and the phonon absorption dominates the phonon emission. But using the equilibrium distribution centered at $\mathbf{k}$=0 to describe the parameters of electron gas moving with average drift velocity $\mathbf{v}_d$ cannot be a good fit with reality, especially when the value of $\mathbf{v}_d$ becomes comparable to the Fermi velocity. Therefore it is more prudent to use the drifted FD distribution that is shifted in the k-space by $\mathbf{k}_d \equiv \mathbf{v}_d$, where the drift velocity is normalized to the value defined above $v_{max} = \hbar q_0/m_c = 4 \times 10^7 cm/s$. The drifted FD distribution

$$f(\mathbf{k},\mathbf{v}_d) = f_0(\mathbf{k}-\mathbf{v}_d) = [\exp(((\mathbf{k}-\mathbf{v}_d)^2 - \mu)/T_e) - 1]^{-1} \qquad (1)$$

is shown in Fig.1b. For small $v_d$ the stimulated emission term remains negative, but for larger values (as shown in Fig 1b), for sufficiently large phonon wavevectors the condition $f(\mathbf{k}_1) > f(\mathbf{k}_2)$ can be reached. This condition can of course be described as population inversion between the two states, and, once it is fulfilled for a large number of pairs of states $\mathbf{k}, \mathbf{k}-\mathbf{q}$, the total stimulated emission rate for the population of LO phonons

$$R_{q,st} = 2\sum_{\mathbf{k}} 2\pi\hbar^{-1} M^2_{\mathbf{k},\mathbf{k}-\mathbf{q}} \delta(E_\mathbf{k} - E_{\mathbf{k}-\mathbf{q}} - 1)/\hbar\omega_{LO} \times [f(\mathbf{k}) - f(\mathbf{k}-\mathbf{q})] n_q = G_q n_q \qquad (2)$$

becomes positive, indicating that the phonons experience gain $G_q > 0$. It is obvious that the LO phonons whose wave-vector is parallel to $\mathbf{v}_d$ will experience the largest population inversion and gain. For these phonons, choosing the polar axis along $v_d$ and introducing the shifted wavevector $\mathbf{k}' = \mathbf{k} - \mathbf{v}_d$ allows one to re-write the delta function in (2) as $\delta\left(\cos\theta - (k'_{min} + v_d)/(k' + v_d)\right)/(k' + v_d)q$, where $\theta$ is the polar angle of $\mathbf{k}'$ and $k'_{min} = \frac{1}{2}(q + q^{-1}) - v_d$. Going from summation to integration one then obtains for the temporal gain

$$G_q = G_0 q^{-3} \int_{k_{min}}^{\infty} [f_k - f_{k-q}](k^2/k + v_d) dk \qquad \text{where we introduce} \qquad G_0 = e^2 q_0 / 4\pi\varepsilon'\hbar = \alpha_0(\varepsilon_0/\varepsilon')q_0 c = 1.2 \times 10^{14} s^{-1}$$

and have dropped the prime symbol in the integral. In new notation $f_k = [\exp((k^2 - \mu)/T_e) + 1]^{-1}$ and $f_{k-q} = [\exp((k^2 + q^2 - 2kq(k_{min} + v_d)/(k + v_d) - \mu)/T_e) + 1]^{-1}$; hence the population inversion condition $f_k > f_{k-q}$ is $\frac{1}{2}(q + q^{-1}) - v_d < k < q^2 v_d$, indicating that it can occur only when $q > 1/2 v_d$. Thus it is expected that, for small $v_d$, only LO phonons with large momentum can experience gain, and this gain is small, but as $v_d$ increases, the gain spectrum spreads towards smaller $q$'s and its value increases. The gain obviously increases with the electron density as the relation between chemical potential $\mu$ and $n_e$ (normalized to the density of electrons with Fermi wave-vector equal to $q_0$, $n_0 = \frac{1}{3\pi^2} q_0^3 = 1.1 \times 10^{19} cm^{-3}$) is $n_e = 3\int [\exp((k^2 - \mu)/T_e) + 1]^{-1} k^2 dk$.

The results of calculating $G_q$ for two different 3-D densities of electrons $N_e = 5 \times 10^{18}$ cm$^{-3}$ ($n_e = 0.45$) and $N_e = 3 \times 10^{19}$ cm$^{-3}$ ($n_e = 2.7$) are shown in Fig.2 (a) and (b) respectively. In both cases the gain, confined to the relatively narrow region of phonon wavevectors, becomes noticeable at drift velocities approaching $2\times10^6$ cm/s ($v_d$~0.2 relative to $v_{max}$). As drift velocity increases both the width and peak value of gain increase rapidly, as the peak slowly shifts towards smaller $q$'s. With higher electron density the gain increases faster as can be seen in Fig.3, where the value of peak phonon gain is plotted as a function of drift velocity for four different carrier densities, assuming $T_e$ of 1030K (normalized $T_e = 1$).

The rate equation for the phonons can be written as

$$\frac{dn_q}{dt} = G_q n_q + S_q - \frac{n_q - \bar{n}_q}{\tau_p} \qquad (3)$$

where $\bar{n}_q$ is the equilibrium phonon occupational number at a given lattice temperature, and the rate of spontaneous emission is $S_q = G_0 q^{-3} \int_{k_{min}}^{\infty} f_k [1 - f_{k-q}](k^2/k + v_d) dk$. Clearly when the peak gain in (3) approaches its threshold value $\tau_p^{-1}$ the number of LO phonons starts increasing exponentially. Therefore, as the bias increases all the additional power transferred from the field to the electrons will be very efficiently transferred to the LO phonons via a stimulated emission process, and the drift velocity will be clamped at the threshold value exactly as the population inversion in the laser is clamped at the threshold [23]; hence it is tempting to refer to this process as "phonon lasing" [21,22]. The major difference from an optical laser is of course the absence of the resonant cavity providing feedback and selecting a single resonant mode. But feedback is not necessary since LO phonons have small group velocities (even when coupled with plasmons). And mode selection is achieved by the sharp spectrum of the gain itself. Whether to call the process "lasing" or "amplified spontaneous emission" is an interesting question from the fundamental physics point of view, but it is not directly relevant to this work in which, far from playing a constructive role, the "phonon lasing" acts as a "spoiler" that limits drift velocity and hence negatively affects both the speed and transconductance of HEMTs. If one is to avoid any reference to "lasing" perhaps the term "phonon avalanche" would be the most appropriate.

If we now plot the line corresponding to the threshold gain in Fig. 2 and 3, we can find the values at which drift velocity saturates for different carrier densities. The value of peak gain curve in Fig.3 increases with carrier densities. therefore the saturation velocity exhibits a strong dependence on electron density (as well as on phonon lifetime) but at really high densities (in excess of $2\times10^{19}$

cm$^{-3}$) the drift velocity is expected to saturate around $10^7$cm/s for a wide range of conditions. This can be best seen in Fig.4a where the density dependence of saturation velocity is compared with the experimental values obtained in [6]. Since our model is three-dimensional, the values of $N_e$ were obtained from the measured sheet density $N_s$ by a self-consistent solution of the Poisson and Schrödinger equations. As one can see, our simple model, using a single adjustable parameter ($T_e \sim$ 1000K), achieves good agreement with the experimental data. About 10% deviation between $N_e$=1 and $2\times10^{19}$ cm$^3$ can probably be explained by the enhanced scattering of LO phonons by hot electrons which effectively broadens the gain and delays the onset of "lasing".

We have also performed modeling using two-dimensional electron gas and three-dimensional electron gas in the channel. Since the phonons remain unconfined and hence three-dimensional, calculations required the effective channel thickness to determine the volume in which the phonons interacting with the electrons are contained. These thicknesses have also been determined by self-consistent solutions to Poisson and Schrödinger equations. The derivation of the expression for the gain is similar to the one outlined above for the 3-D case, and the final result is

$$G_q = 2G_0 \frac{1}{q^3 d_{eff}} \int_{k_{min}}^{\infty} [f_k - f_{k-q}] \frac{k}{\sqrt{(k+v_d)^2 - (k_{min}+v_d)^2}} dk \qquad (4)$$

where the effective thickness of channel $d_{eff}$ is scaled by $q_0$. The results of the calculation are shown in Fig.4b and they reproduce the experimental data reasonably well, yet not as well as 3D modeling. That may be due to the fact that the calculations do not take into account the exact shape of electron wavefunction – just its effective thickness.

Having shown that newly-introduced "phonon lasing" or "phonon avalanche" model predicts the concentration dependence of $v_{sat}$ with a good accuracy, it is interesting to explore the dynamics of phonons and carriers at different electric fields without resorting to the complexity of full Monte-Carlo simulations. This can be accomplished by solving balance equations for kinetic energy and average velocity of electrons, namely

$$\begin{aligned}\frac{dE_k}{dt} &= -Fv_d - (3/8\pi n_r)\int \left[n_q(T_e) - \overline{n}_q\right] d^3\mathbf{q} = 0 \\ \frac{dv_d}{dt} &= -F/2 - (3/8\pi n_r)\int \left[n_q(T_e) - \overline{n}_q\right] q_x d^3\mathbf{q} - \mu_{LF}^{-1} v_d = 0\end{aligned} \qquad (5)$$

where kinetic energy $E_k$ is normalized to $\hbar\omega_{LO}$, time to $\tau_p$, the electric field $F$ to $F_0 = \hbar q_0 / 2e\tau_p = 1.2\times 10^5 V/m$, and the low field mobility (which includes all the scattering mechanisms other than LO phonons) $\mu_{LF}$ to $\mu_0 = e\tau_p/m_c = 2.3\times 10^5 cm^2/V\cdot s$ With the temperature-dependent $n_q$ obtained from (3), for each value of $v_d$, a pair of $F$ and $T_e$ satisfying (5) can be obtained, resulting in the velocity saturation curves shown in Fig.5a for the same $N_e = 5\times 10^{18}$ cm$^{-3}$ ($n_e = 0.45$) and $N_e = 3\times 10^{19}$ cm$^{-3}$ ($n_e = 2.7$) as in Fig.2. The lattice temperature of 350K and low-field mobility of 1000 cm$^2$/V-s has been assumed, and the results conform both to the estimates of $v_{\text{sat}}$ made above (Fig.4a) and with the experimental saturation curves measured in [6]. To illustrate the role played by the stimulated emission of LO phonons, five "snapshots" of LO phonon distribution in the BZ at different drift velocities (as indicated in Fig.5a) are displayed in Fig.5b and the values of $n_q$ vs. $q_x$ are plotted in Fig.5c. As one can see, for small values of $v_d$, the phonon distribution is spread out in the BZ, but as $v_d$ increases, a clearly pronounced narrow peak appears on the $q_x$ axis in which the $n_q$ grows exponentially and $v_d$ saturates.

The shortcoming of this simple model is that it fails to predict what exactly happens beyond the saturation point. In this respect the situation no different from the fact that the behavior of a silicon field effect transistor (FET) beyond saturation (pinch-off) requires a more involved description than a textbook FET model [24]. Nevertheless, this model is quite useful as it faithfully predicts saturation voltage and current. Similarly, our model based on the shifted FD distribution characterized by a single $T_e$ faithfully predicts the behavior of GaN HFET up to the point where the stimulated emission of LO phonons inside a narrow volume region in reciprocal space becomes the dominant relaxation process. At this point, the electron-electron scattering is no longer fast enough to replenish the population of forward propagating electrons that is quickly depleted via stimulated LO emission; hence the carrier distribution is no longer in equilibrium. The situation is strikingly similar to the "spectral hole burning" [25] occurring in lasers (the "spectrum" here is of course in the spatial rather than the frequency domain). As a result, the distribution of LO phonons (Fig.5b) is expected to saturate and broaden, and the drift velocity is expected to continue to increase slowly, as indicated by the dashed lines in Fig.5a and had been indeed measured in [6].

Before concluding, we would like to assess the feasibility of experimental confirmation of "phonon lasing" that is more direct than successful reproduction of the experimental curve. Since LO phonons have very small velocities, they remain confined inside the HEMT channel; hence the best means to reveal their existence in the small region of BZ is by the second order Anti-Stokes Raman measurements of the operating HEMT. These experiments are far from trivial but are within realm of possible. [26]. Frankly, we can see no practical applications of these hot phonons as yet, unless the decay of the stationary LO phonons can generate directional acoustic THz phonons. This issue should be investigated in future work. The main practical consequence of this work is that "lasing-like" stimulated emission of LO phonons has been identified as a major force causing saturation of the drift velocity in GaN HEMT. Controlling, or, better preventing the "phonon lasing" by introducing additional LO phonon scattering mechanisms like disorder [27] may lead to improvement of III-nitride HEMT's performance.

Acknowledgement

The authors gratefully acknowledge the support from DATE MURI (ONR N00014-11-1-0721, Dr. Paul Maki)

References

1. Y. Yue, Z. Hu, J. Guo, B. Sensale-Rodriguez, G. Li, R. Wang, F. Faria, T. Fang, B. Song, X. Gao, S. Guo, T. Kosel, G. Snider, P. Fay, D. Jena & H. Xing, IEEE Electron Device Lett. **33**(7), 988-990 (2012)

2. K. Shinohara, D. Regan, A. Corrion, D. Brown, Y. Tang, J. Wong, G. Candia, A. Schmitz, H. Fung, S. Kim and M. Micovic, IEEE IEDM, **27**.2.1-27.2.4 (2012)

3. Y. Tang; K. Shinohara, D. Regan, A. Corrion, D. Brown, J. Wong, A. Schmitz, H. Fung, S. Kim, M. Micovic, IEEE Electron Device Lett. **36**(6), 549-551 (2015)

4. S. Dasgupta, D. F. Brown, F. Wu, S. Keller, J. S. Speck, & U. K. Mishra, *Appl. Phys. Lett.* **96**(14), 143504 (2010).

5. D. K. Ferry, "Semiconductor Transport", Taylor and Francis, London, (2001)

6. S. Bajaj, O. F. Shoron, P. S.Park, S. Krishnamoorthy, F.Akyol, T-H Hung, Sh. Reza, E. Chumbes, J. Khurgin and S. Rajan, Appl. Physics Lett. 107 ,153504 (2015)

7. B. K. Ridley, W. J. Schaff, and L. F. Eastman, J. Appl. Phys. **96**(3), 1499-1502 (2004)

8. J. Khurgin, Y. Ding & D. Jena, Appl. Phys. Lett. **91**(25), 252104 (2007)

9. J. H. Leach, C. Y. Zhu, M. Wu, X. Ni, X. Li, J. Xie, Ü. Özgür, Hadis Morkoç, J. Liberis, E. Šermukšnis, A. Matulionis, H. Cheng, and Ç. Kurdak , Appl. Phys. Lett, **95**, 223504 (2009)

10. G. P. Srivastava Phys. Rev. B 77, 155205 (2008)

11. G. P. Srivastava, J. Phys.: Condens. Matter 21 174205 (2009)
12. K. T. Tsen, D. K. Ferry, A. Botchkarev, A. Serdlov, A. Salvador, and H. Morkoc, Appl. Phys. Lett. **72**, 2132 (1998)
13. L. Shi, F. A. Ponce, and J. Menendez, Appl. Phys. Lett. **84**, 3471 (2004).
14. P. G. Klemens, Phys. Rev. **148**, 845 (1966)
15. B. K. Ridley, J. Phys. Condens. Matter **8**, L511 (1996).
16. K. T. Tsen, J. G. Kiang, D. K. Ferry, and H. Morkoc, Appl. Phys. Lett. 89, 112111 (2006).
17. A. Matulionis, J. Liberis, I. Matulionienė, M. Ramonas, L. F. Eastman, J. R. Shealy, V. Tilak, and A. Vertiatchikh Phys. Rev. B **68**, 035338 (2003)
18. A. Matulionis, J. Liberis, E. Šermukšnis, J. Xie, J. H. Leach, M. Wu, and H. Morkoç, Semicond. Sci. Technol. **23**, 075048 (2008).
19. J. B. Khurgin, S. Bajaj, and S. Rajan, , Appl. Phys. Lett., **107**, 262101 (2015)
20. M. Ramonas and A Matulionis Monte Carlo simulation of *Semicond. Sci. Technol.* **19** S424 (2004)
21. R. P. Beardsley, A. V. Akimov, M. Henini, and A. J. Kent, Phys. Rev. Lett. **104,** 085501 (2010)
22. I. S. Grudinin, H. Lee, O. Painter, and K. J. Vahala, Phys. Rev. Lett **104**, 083901 (2010)
23. A. E. Siegman, "Lasers", University Science Books, Sausalito Ca (1996)
24. D. A. Neamen, "Semiconductor Physics and Devices", McGrow Hill, NY (2012)
25. O. Svelto, D. C. Hanna, "Principles of Lasers", Plenum, NY (2010)
26. G. Xu, S. K. Tripathy, Y. J. Ding, Y. Cao, K. Wang, D. Jena, J. B. Khurgin, Laser Physics, **19**, 745 (2009)
27. J.B. Khurgin, D. Jena, YJ Ding, Appl. Phys. Lett., **93** 032110 (2008)

**Figure Captions**

**Fig.1** Electron distribution in the BZ using the standard **(a)** and drifted **(b)** FD functions, which results in absorption (**a**) or gain (**b**) for LO phonons.

**Fig.2** Wave vector dependence of the LO phonon gain $G_q$ for two different electron densities **(a)** $N_e=5\times10^{18}$ cm$^{-3}$ and **(b)** $N_e=5\times10^{18}$ cm$^{-3}$

**Fig.3** Peak value of LO phonon gain $G_q$ as a function of drift velocity $v_d$ for different electron densities.

**Fig.4** Saturation velocity in GaN HEMT as a function of electron density. (a) Three dimensional model (b) Two dimensional model. Red dots represent experimental data from [6].

**Fig.5** (a) Velocity vs. field curves of GaN HEMT for two electron densities modeled up to the onset of phonon "lasing" or ("avalanche") (b,c) Hot LO phonon distribution in the BZ zone for 5 regimes indicated in (a).

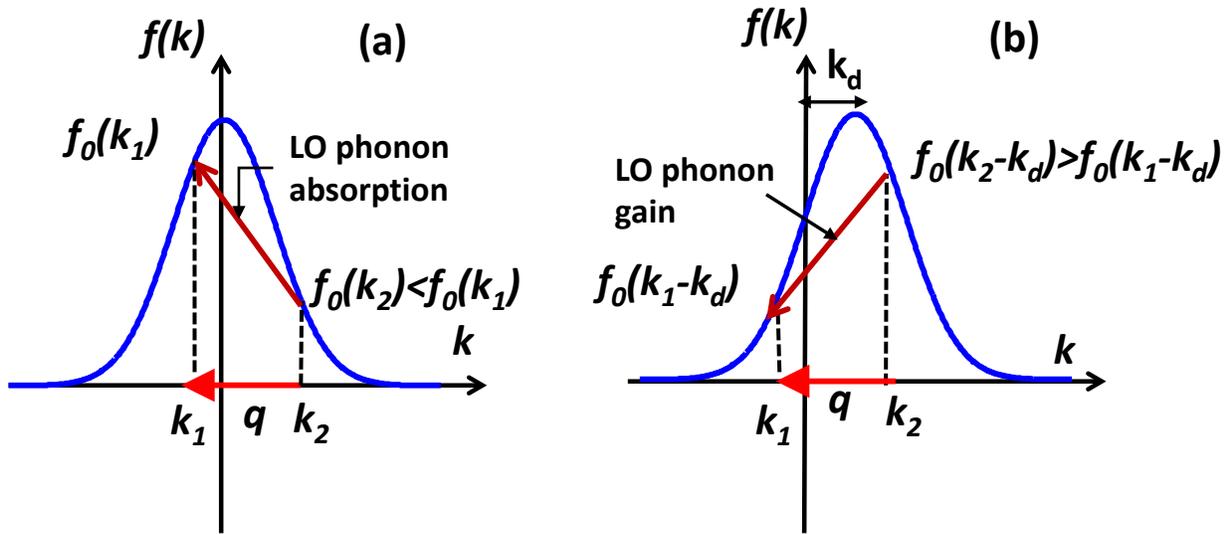

**Fig.1** Electron distribution in the Brillouin zone using the standard (**a**) and drifted (**b**) FD functions, which results in absorption (**a**) or gain (**b**) for LO phonons.

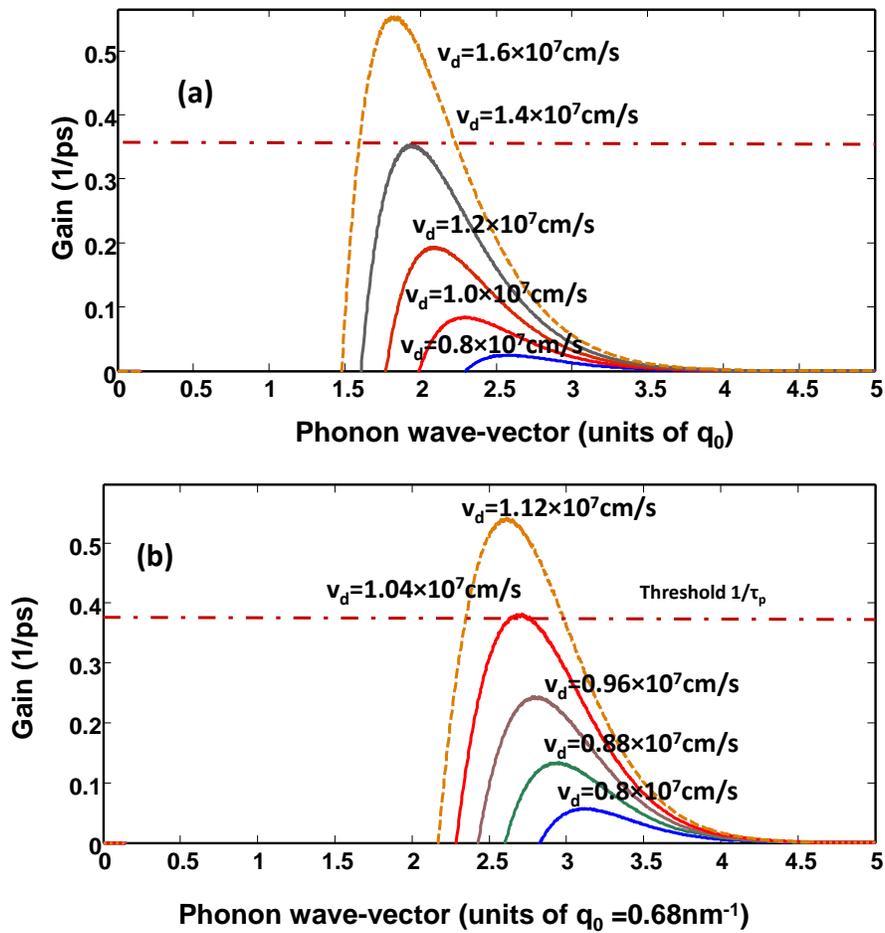

**Fig.2** Wave vector dependence of the LO phonon gain $G_q$ for two different electron densities (**a**) $N_e=5\times10^{18}$ cm$^{-3}$ and (**b**) $N_e=5\times10^{18}$ cm$^{-3}$

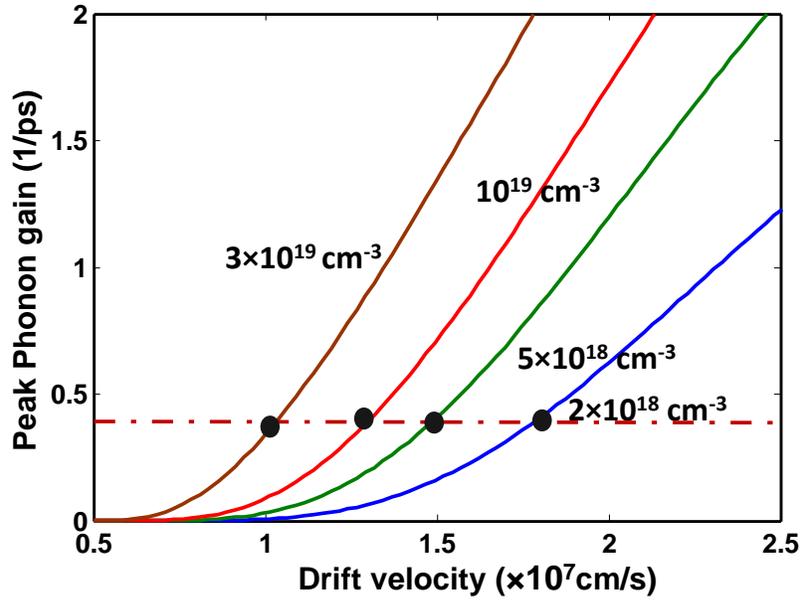

**Fig.3** Peak value of LO phonon gain $G_q$ as a function of drift velocity $v_d$ for different electron densities.

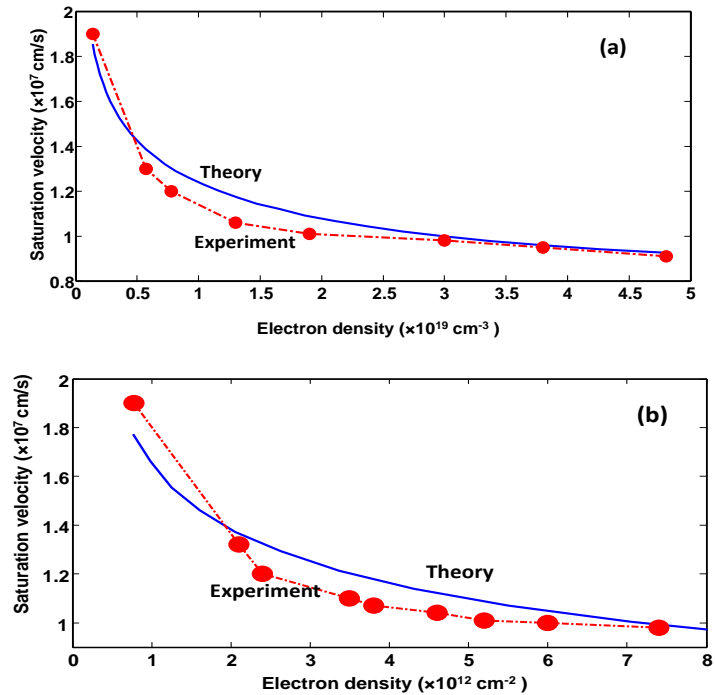

**Fig.4** Saturation velocity in GaN HEMT as a function of electron density. (a) Three dimensional model (b) Two dimensional model. Red dots represent experimental data from Ref. [6]

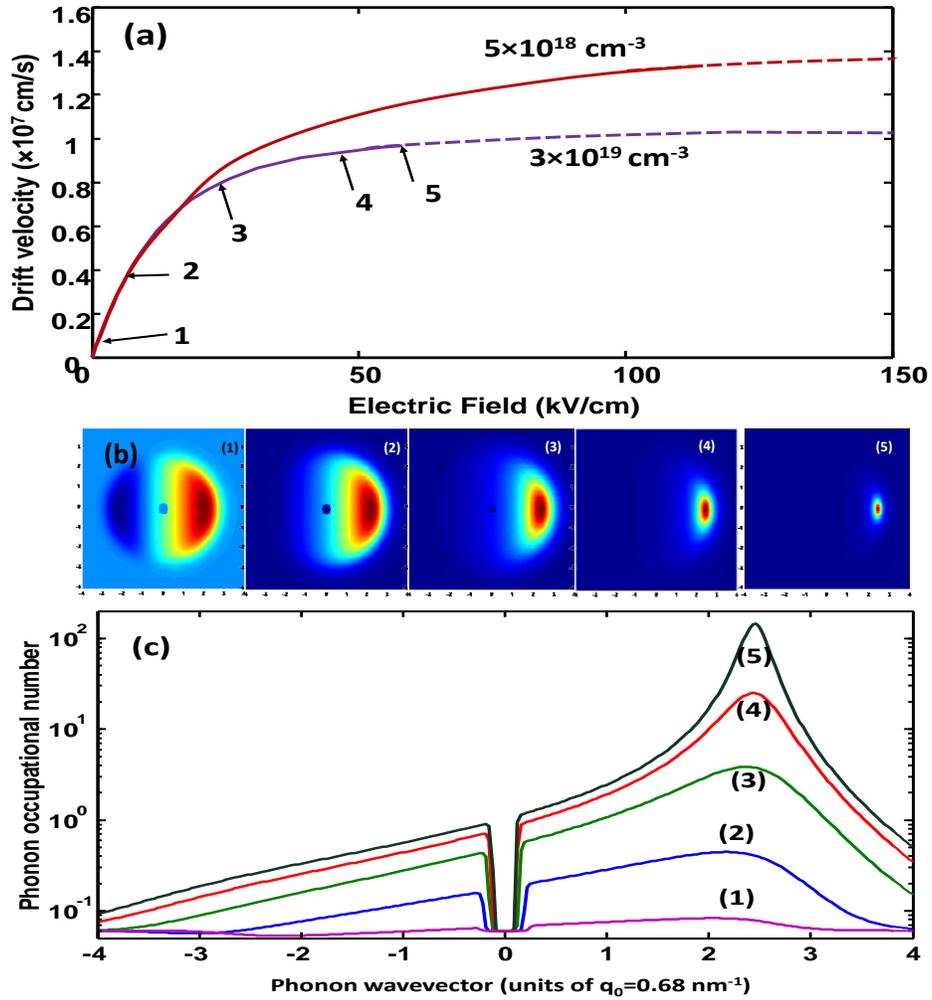

**Fig.5** (a) Velocity vs. field curves of GaN HEMT for two electron densities modeled up to the onset of phonon "lasing" or ("avalanche") (b,c) Hot LO phonon distribution in the Brillouin zone for 5 regimes indicated in (a).